\documentstyle[amsfonts,aps]{revtex}

\begin{document}
\title{$SU_L\left( 4\right) \times U\left( 1\right) $ model for electroweak
unification}
\author{Fayyazuddin and Riazuddin}
\address{National Center for Physics\\
Quaid-i-Azam University\\
Islamabad 45320\\
Pakistan}
\date{NCP-QAU/2004-003, hep-ph/0403042}
\maketitle

\begin{abstract}
After some general remarks about $SU_L\left( 4\right) $ electroweak
unification, the model is extended to $SU_L\left( 4\right) \times U_X\left(
1\right) $ to accomodate fractionally charged quarks. The unification scale
is expected to be in TeV region. A right-handed Majorana neutrino along with
known lepton are put in the fundamental representation of $SU_L\left(
4\right) $ with $Y_X=0$. The see-saw mechanism for neutrino masses and
flavor mixing in neutrino sector is a natural feature of the model. The
lepton number violating processes can occure through dilepton gauge bosons
contained in the model.
\end{abstract}

\section{Introduction}

It is well-known that the standard $SU_L\left( 2\right) \times U\left(
1\right) $ model does not provide a true unification of electroweak
interaction, since because of the $U\left( 1\right) $ factor, $\sin ^2\theta
_W$ is not fixed by the model. The extensions of the standard model to
higher groups such as $SU\left( 5\right) $, $O\left( 10\right) $ or
Pati-Salam group $SU_c\left( 4\right) \times SU_L\times S_R\left( 2\right) $
do give quantisation of charge and predict $\sin ^2\theta _W=3/8$ at
unification scale; this unification scale is of order 10$^{16}$ GeV \cite
{r01,r02,r03}. The reason for the huge desert is the large disparity between
the measured value of $\sin ^2\theta _W=0.231$ and $\sin ^2\theta _W=3/8$ at
the unification scale. The most attractive model in this catagory of
unification is the supersymmetric standard model based on the group $%
SU\left( 5\right) $ (SSM) \cite{r04}.

There are other models based on the simple groups $SU\left( 3\right) $ or $%
SU\left( 4\right) $, which gives $\sin ^2\theta _W=1/4$ at the unification
scale. For the group $SU_L\left( 3\right) $ \cite{r05,r06,r07}, the
quantised charge $Q$ can be written as 
\begin{equation}
Q=\frac 12\left[ \lambda _3+\sqrt{3}\lambda _8\right] =T_3+\sqrt{3}T_8=T_3+%
\frac 12Y  \label{e01}
\end{equation}
Thus leptons can be assigned naturally to the representation 
\begin{equation}
3_R=\left( 
\begin{tabular}{c}
$e^{+}$ \\ 
$\nu _e^c$ \\ 
$e^{-}$%
\end{tabular}
\right) _R\Longrightarrow 3_L^{*}=\left( 
\begin{tabular}{c}
$e^{-}$ \\ 
$\nu _e$ \\ 
$e^{+}$%
\end{tabular}
\right) _L  \label{e02}
\end{equation}
of $SU_L\left( 3\right) $. For this case 
\begin{eqnarray}
\frac{A_\mu }e &=&\frac{W_{3L\mu }}g+\frac{B_\mu }g,  \label{e03} \\
\frac 1{e^2} &=&\frac 1{g^2}+\frac 1{g^{\prime 2}}  \nonumber \\
\tan \theta _W &=&\frac{g^{\prime }}g  \label{e04}
\end{eqnarray}
Since in the symmetry limit $g^{\prime }=g/\sqrt{3}$, it gives at the
unification scale $\sin ^2\theta _W=1/4=0.25$ near to the measured value and
thus unification scale is expected to be in TeV region. This is an
attractive feature of this group. However, quarks cannot be accomodated in
this model. It needs to be extended to $SU_L\left( 3\right) \times U_X\left(
1\right) $ in order to accomodate the quarks. The anomely free model based
on the group $SU_L\left( 3\right) \times U_X\left( 1\right) $ has been
considered by several authors \cite{r06,r07,r08}.

A unification model based on the group $SU_L\left( 4\right) $ has been
considered in Ref. \cite{r09}. It not only accomodates leptons nicely; but
also a right-handed Majorana neutrino. The see-saw mechanism for generation
of masses of the standard model neutrinos is a natural feature of this
model. The charge operator in this model is given by 
\begin{equation}
Q=\frac 12\left[ \lambda _3+\frac 1{\sqrt{3}}\lambda _8\right] +\sqrt{\frac 2%
3}\lambda _{15}  \label{e06}
\end{equation}
The leptons can be assigned to the fundamental representation of this group
(Ist generation) as follows: 
\begin{equation}
F_e=\left( 
\begin{array}{c}
e^{+} \\ 
\nu _e^c \\ 
-N_e \\ 
e^{-}
\end{array}
\right) _R:4_R  \label{e07}
\end{equation}
It is convienent to write the charge operator in the form 
\begin{equation}
Q=\frac 12\left[ \tau _{3L}+\tau _{3R}+Y_1\right]  \label{e08}
\end{equation}
where 
\begin{eqnarray}
\lambda _3 &=&\left( 
\begin{tabular}{cc}
$\tau _{3L}$ & 0 \\ 
0 & 0
\end{tabular}
\right) ,  \nonumber \\
\frac 1{\sqrt{3}}\left( -\lambda _8+\sqrt{2}\lambda _{15}\right) &=&\left( 
\begin{tabular}{cc}
0 & 0 \\ 
0 & $\tau _{3R}$%
\end{tabular}
\right)  \label{e09}
\end{eqnarray}

and 
\begin{equation}
Y_1=\frac 1{\sqrt{3}}\left( 2\lambda _8+\sqrt{2}\lambda _{15}\right)
=diag\left( 1,1,-1,-1\right)  \label{e10}
\end{equation}
The vector bosons belong to the adjoint representation of $SU_L\left(
4\right) $: (suppressing the Lorentz index $\mu $) 
\[
W=\frac 1{\sqrt{2}}\left( 
\begin{tabular}{cccc}
$W_{3L}+\frac 1{\sqrt{2}}B_1$ & $\sqrt{2}W_L^{-}$ & $\sqrt{2}X_1^{-}$ & $%
\sqrt{2}Y_1^{--}$ \\ 
$\sqrt{2}W_L^{+}$ & -$W_{3L}+\frac 1{\sqrt{2}}B_1$ & $\sqrt{2}X_2^0$ & $%
\sqrt{2}Y_2^{-}$ \\ 
$\sqrt{2}X_1^{+}$ & $\sqrt{2}\bar{X}_2^0$ & $W_{3R}-\frac 1{\sqrt{2}}B_1$ & $%
\sqrt{2}W_R^{-}$ \\ 
$\sqrt{2}Y_1^{++}$ & $\sqrt{2}Y_2^{+}$ & $\sqrt{2}W_R^{+}$ & $-W_{3R}-\frac 1%
{\sqrt{2}}B_1$%
\end{tabular}
\right) 
\]
Thus 
\begin{eqnarray}
\frac{A_\mu }e &=&\frac{W_{3L\mu }}g+\frac{W_{3R\mu }}g+\frac{B_{1\mu }}{g_1}%
=\frac{W_{3L\mu }}g+\frac{B_\mu }{g^{\prime }}  \label{e12} \\
\frac{B_\mu }{g^{\prime }} &=&\frac{W_{3R\mu }}g+\frac{B_{1\mu }}{g_1}
\label{e13} \\
\frac 1{e^2} &=&\frac 2{g^2}+\frac 1{g^{\prime 2}}  \label{e14} \\
\frac 1{g^{\prime 2}} &=&\frac 1{g^2}+\frac 1{g_1^2}  \label{e15}
\end{eqnarray}
In the symmetry limit $g_1=g/\sqrt{2}$ and hence at the unification scale $%
g_1=g/\sqrt{3}$, and $\sin ^2\theta _W=1/4$.

Before we proceed further, it is interesting to see that in $SU_L\left(
4\right) $, there is an alternative way to write the charge operator: 
\begin{eqnarray}
Q &=&\frac 12\left[ \lambda _3-\frac 1{3\sqrt{3}}\lambda _8+\frac 43\sqrt{%
\frac 23}\lambda _{15}\right]  \nonumber \\
&=&\frac 12\left[ \tau _{3L}+\tau _{3R}+Y_1\right]  \label{e16}
\end{eqnarray}
where 
\begin{equation}
Y_1=\frac 1{3\sqrt{3}}\left( 2\lambda _8+\sqrt{2}\lambda _{15}\right)
\label{e17}
\end{equation}
In this case $Y_1=diag\left( 1/3,1/3,-1/3,-1/3\right) $. It is then natural
to assign the quarks to the fundamental representation of $SU_L\left(
4\right) $ as follows 
\begin{equation}
F_q=\left( 
\begin{array}{c}
u \\ 
d \\ 
d^c \\ 
u^c
\end{array}
\right) :4_L  \label{e18}
\end{equation}
However, for this case, at the unification scale $\sin ^2\theta _W=9/20=0.45$%
; hence the unification scale may be as high as Planck mass. $\Delta B=2$
transition will be highly suppressed. We conclude that for $SU_L\left(
4\right) $, the natural choice for charge operator is that given in Eq. (\ref
{e04}); the leptons belong to the fundamental representation of $SU_L\left(
4\right) $, the unification scale is in TeV region; the lepton number
violation occures at TeV scale; Majorana mass for the right-handed neutrino
is also large and see-saw mechanism is a natural consequence of $SU_L\left(
4\right) $.

We note that representation (\ref{e07}) is not an anomaly free. An anomaly
free model can be constructed by putting Hahn and Nambu quarks \cite{r10}
with integral gauge charges in the fundamental representation $SU_L\left(
4\right) $ as follows (for first generation) 
\begin{eqnarray}
F_d^r &=&\left( 
\begin{array}{c}
d^{\prime c} \\ 
u^c \\ 
-u \\ 
d^{\prime }
\end{array}
\right) _R^r:4_R\Longrightarrow 4_L^{*}  \nonumber \\
F_d^{g,b} &=&\left( 
\begin{array}{c}
u \\ 
-d \\ 
d^c \\ 
u^c
\end{array}
\right) _L^{g,b}:4_L  \label{e19}
\end{eqnarray}
It is clear that representations (\ref{e07}) and (\ref{e19}) together cancel
the anomalies. The gauge symmetry $SU_L\left( 4\right) $ is broken as
follows 
\begin{equation}
SU_L\left( 4\right) 
\begin{tabular}[t]{c}
$\rightarrow $ \\ 
$m_{X,Y}$%
\end{tabular}
SU_L\left( 2\right) \times SU_R\left( 2\right) \times U_1\left( 1\right) 
\begin{tabular}[t]{c}
$\rightarrow $ \\ 
$m_R$%
\end{tabular}
SU_L\left( 2\right) \times U_1\left( 1\right) 
\begin{tabular}[t]{c}
$\rightarrow $ \\ 
$m_L$%
\end{tabular}
U_{em}\left( 1\right)  \label{e20}
\end{equation}
As we have emphasized earlier this model works well for leptons. For quarks,
the following comments are in order. The electromagnetic current can be
split into two parts viz, the color singlet (containing the fractionally
charged quarks) and a color octet. The currents coupled to $X$ and $Y$
bosons do not contain any color singlets. Thus with the singlet part of the
electromagnet current; it effectively reduces to the standard model of
electroweak interactions \cite{r09}, since known hadrons are color singlets.
However, this model becomes more appealing if we could formulate this model
in higher dimensions and gauge symmetry is broken by orbifold
compactification to four dimensions so that $X$ and $Y$ bosons and non-color
singlet part of electromagnetic current is relagated to the bulk and we have
effectively fractional quarks on the brane.

\section{Symmetry group $SU_L\left( 4\right) \times U_X\left( 1\right) $}

In order to accomodate the quarks with fractional charges, it is necessary
to extend the group to $SU_L\left( 4\right) \times U_X\left( 1\right) $.

However, for the extended group $SU_L\left( 4\right) \times U_X\left(
1\right) $, true unification is lost and $\sin ^2\theta _W$ is arbitrary.
Neverthless for this group, $\sin ^2\theta _W$ comes to be less than 1/4 and
it is still possible to have unification mass scale in TeV region. To see
this we note that for the extended group we have an extra gauge boson $V_\mu
\left( x\right) $ with gauge coupling constant $g_x$. In the limit $%
g_x\rightarrow \infty $, the vector boson $V_\mu \left( x\right) $ is
decoupled, we have effectively the group $SU_L\left( 4\right) $, with the
unification mass scale $m_X\sim 5.8$ TeV and $\sin ^2\theta _W\left(
m_X\right) =1/4$. However with finite $g_X$, we have $\sin ^2\theta _W\left(
m_X\right) <1/4$, $m_X<5.8$ TeV. The lower bound on unification mass scale
would be determined by the experimental limit on lepton number violating
processes. However, the smallness of neutrino masses, would require the
Majorana mass of right-handed neutrino to be in TeV region. Thus one would
expect the unification mass scale $m_X$ to be few TeV. In fact we find for $%
\alpha _2\left( m_Z\right) /\alpha _X\left( m_Z\right) \approx 0.147$, $%
m_X=960$ GeV, where as for $\alpha _2\left( m_Z\right) /\alpha _X\left(
m_Z\right) \approx 0.052$, $m_X=3$ TeV.

As long as $g_X\left( m_Z\right) /g_2\left( m_Z\right) \gg 1$, it is still
possible to have a sort of unification at a TeV mass scale (For example, for 
$g_X\left( m_Z\right) /g_2\left( m_Z\right) \approx 4.4$, $m_X=3$ TeV).

We now give the details of the model. For anomely cancelation, the leptons
and quarks must be assigned as follows: 
\begin{eqnarray}
&&\left. 4_R:\left( 
\begin{array}{c}
e^{+} \\ 
-\nu _e^c \\ 
N_e^{\prime } \\ 
e^{-}
\end{array}
\right) _R,\,\,\,\left( 
\begin{array}{c}
\mu ^{+} \\ 
-\nu _\mu ^c \\ 
N_\mu ^{\prime } \\ 
\mu ^{-}
\end{array}
\right) _R,\,\,\,\left( 
\begin{array}{c}
\tau ^{+} \\ 
-\nu _\tau ^c \\ 
N_\tau ^{\prime } \\ 
\tau ^{-}
\end{array}
\right) _R\Rightarrow 4_L^{*}\,\,\,\,\,\,\,\,\,\,\,\,\,Y_X=0\right. 
\nonumber \\
&&\left. 4_L:\left( 
\begin{array}{c}
u^a \\ 
d^{\prime a} \\ 
D^a \\ 
H_d^a
\end{array}
\right) _L,\,\,\,\,\left( 
\begin{array}{c}
c^a \\ 
s^{\prime a} \\ 
S^a \\ 
H_s^a
\end{array}
\right) _L,\,\,\,\,\,\,\,\,\,\,\,\,Y_X=-2/3\right.  \nonumber \\
&& 
\begin{array}{c}
I_R: \\ 
Y_X=
\end{array}
\begin{array}{c}
\left( u^a\right) _R, \\ 
4/3
\end{array}
\begin{array}{c}
\left( d^{\prime a}\right) _R \\ 
-2/3
\end{array}
\begin{array}{c}
\left( D^a\right) _R \\ 
-2/3
\end{array}
\begin{array}{c}
\left( H_d^a\right) _R \\ 
-8/3
\end{array}
\begin{array}{c}
,
\end{array}
\begin{array}{c}
\left( c^a\right) _R, \\ 
4/3
\end{array}
\begin{array}{c}
\left( s^{\prime a}\right) _R \\ 
-2/3
\end{array}
\begin{array}{c}
\left( S^a\right) _R \\ 
-2/3
\end{array}
\begin{array}{c}
\left( H_s^a\right) _R \\ 
-8/3
\end{array}
\nonumber \\
&&  \label{e26} \\
&&\left. 4_R:\left( 
\begin{array}{c}
b^{\prime a} \\ 
t^a \\ 
U^a \\ 
T^a
\end{array}
\right) _R^c,\,\,\,\,\,\,\,\,\,\,\,\,\,\,\,\,\,Y_X=-4/3\right.  \nonumber \\
&& 
\begin{array}{c}
I_L: \\ 
Y_X=
\end{array}
\begin{array}{c}
\left( b^{\prime a}\right) _L^c, \\ 
+2/3
\end{array}
\begin{array}{c}
\left( t^{\prime a}\right) _L^c, \\ 
-4/3
\end{array}
\begin{array}{c}
\left( U^a\right) _L^c \\ 
-4/3
\end{array}
\begin{array}{c}
\left( T_d^a\right) _L^c \\ 
-10/3
\end{array}
\label{e27}
\end{eqnarray}
Here $a=1,2,3$ is the color index. Thus we have six extra quarks (exotic)
with charges 
\begin{eqnarray}
U,D,S &:&\left( 2/3,-1/3,-1/3\right)  \nonumber \\
H_d,H_s,T &:&\left( -4/3,-4/3,5/3\right)  \label{e28}
\end{eqnarray}

The photon $A_\mu $ and boson $B_\mu $ associated with $U_Y\left( 1\right) $
are given by 
\begin{eqnarray}
\frac{A_\mu }e &=&\frac{W_{3L\mu }}g+\frac{W_{3R\mu }}g+\frac{B_{1\mu }}{g_1}%
+\frac{V_\mu }{g_X}=\frac{W_{3L\mu }}g+\frac{B_\mu }{g^{\prime }}  \nonumber
\label{e12} \\
\frac{B_\mu }{g^{\prime }} &=&\frac{W_{3R\mu }}g+\frac{B_{1\mu }}{g_1}+\frac{%
V_\mu }{g_X}  \label{e22a} \\
\frac 1{e^2} &=&\frac 2{g^2}+\frac 1{g_1^2}+\frac 1{g_X^2}=\frac 2{g^2}+%
\frac 1{g^{\prime 2}}  \nonumber \\
\frac 1{g^{\prime 2}} &=&\frac 1{g^2}+\frac 1{g_1^2}+\frac 1{g_X^2}
\label{e23a}
\end{eqnarray}
In the symmetry limit of $SU_L\left( 4\right) $, $g_1=g/\sqrt{2}$; thus we
get 
\begin{eqnarray}
\frac 1{g^{\prime 2}} &=&\frac 3{g^2}+\frac 1{g_X^2}  \label{e24a} \\
\frac{g^2}{g_X^2} &=&\frac{1-3\tan ^2\theta _W\left( m_X\right) }{\tan
^2\theta _W\left( m_X\right) }  \nonumber \\
\frac{g_X^2}{g^2} &=&\frac{\sin ^2\theta _W\left( m_X\right) }{1-4\sin
^2\theta _W\left( m_X\right) }  \label{e25a}
\end{eqnarray}
Thus $\sin ^2\theta _W\left( m_X\right) <1/4$, implying the unification
scale to be in TeV range. For $\sin ^2\theta _W\left( m_X\right) =1/4$, $%
g_X\rightarrow \infty $ and $V_\mu $ will decouple at $m_X$.

Here it is relevant to give an estimate of the unification scale. We first
consider the unification group $SU_L\left( 4\right) $ and its breaking
direct to the group $SU_L\left( 2\right) \times U_Y\left( 1\right) $. A
straight forward application of renormalization group equations give \cite
{r09} 
\begin{equation}
\alpha ^{-1}\left( m_Z\right) \left[ 1-4\sin ^2\theta _W\right] =2\left(
-C_1^2\beta _2+\beta _1\right) \ln \frac{m_X}{m_Z}  \label{e27b}
\end{equation}
where 
\begin{eqnarray}
\beta _2 &=&\frac 1{4\pi }\left[ -\frac{22}3+\frac 43\frac{n_f}2\right] 
\nonumber \\
\beta _1 &=&\frac 1{4\pi }\left[ \frac 43\frac{n_f}2C_1^2\right] \\
C_1^2 &=&2  \nonumber
\end{eqnarray}
Using $\sin ^2\theta _W\left( m_Z\right) =0.2311$, $\alpha ^{-1}\left(
m_Z\right) =128$, we get unification mass scale 
\begin{equation}
m_X=63.1m_Z\simeq 5.8TeV  \label{a28}
\end{equation}
We now consider the group $SU_L\left( 4\right) \times U_X\left( 1\right) $.
For this case, the renormalization group equations give 
\begin{eqnarray}
\alpha _L^{-1}\left( m_Z\right) &=&\alpha _G^{-1}+2\beta _{2L}\ln \frac{m_X}{%
m_Z}  \nonumber \\
\alpha _R^{-1}\left( m_Z\right) &=&\alpha _G^{-1}+2\beta _{2R}\ln \frac{m_X}{%
m_Z}  \nonumber \\
\alpha _1^{-1}\left( m_Z\right) &=&C_1^2\alpha _G^{-1}+2\beta _1\ln \frac{m_X%
}{m_Z}  \nonumber \\
\alpha _R^{-1}\left( m_Z\right) +\alpha _1^{-1}\left( m_Z\right) &=&\left(
1+C_1^2\right) \alpha _G^{-1}+2\left( \beta _{2R}+\beta _1\right) \ln \frac{%
m_X}{m_Z}
\end{eqnarray}
Noting that $\beta _{2L}=\beta _{2R}=\beta _2$, $\alpha _L^{-1}=\alpha
_2^{-1}$, and $\alpha _2^{-1}\left( m_Z\right) =\alpha ^{-1}\left(
m_Z\right) \sin ^2\theta _W$, we obtain 
\begin{equation}
\alpha _R^{-1}\left( m_Z\right) +\alpha _1^{-1}\left( m_Z\right) =\left(
1+C_1^2\right) \alpha ^{-1}\left( m_Z\right) \sin ^2\theta _W\left(
m_Z\right) +2\left( -C_1^2\beta _2+\beta _1\right) \ln \frac{m_X}{m_Z}
\end{equation}
Noting that 
\begin{equation}
\left. \alpha ^{\prime }\right. ^{-1}=\alpha _R^{-1}+\alpha _1^{-1}+\alpha
_X^{-1},
\end{equation}
we obtain our final result 
\begin{equation}
\frac{\alpha _X^{-1}\left( m_Z\right) }{\alpha _2^{-1}\left( m_Z\right) }=%
\frac{1-4\sin ^2\theta _W}{\sin ^2\theta _W}-2\alpha _2\left( m_Z\right)
\left( -C_1^2\beta _2+\beta _1\right) \ln \frac{m_X}{m_Z}
\end{equation}

This relation reduces to Eq. (\ref{e27b}) for $\alpha _X^{-1}=0$. Thus from
Eq. (\ref{a28}), we conclude that upper bound for the unification mass scale
is 5.8 TeV, for which $\alpha _X^{-1}=0$. The unification scale of order 900
GeV, would correspond to $\alpha _X^{-1}\left( m_Z\right) /\alpha
_2^{-1}\left( m_Z\right) \simeq 0.126$ whereas for the unification mass
scale of 3 TeV, this ratio is 0.052.

We note that there are four neutral vector bosons: $W_{3L\mu }$, $B_{1\mu }$%
, $W_{3R\mu }$, and $V_\mu $. They can be expressed in terms of the photon $%
A_\mu $, the neutral vector boson $Z_\mu $ and two extra neutral bosons $%
Z_\mu ^{\prime }$ and $Z_\mu ^{\prime \prime }$ as follows 
\begin{eqnarray}
A_\mu &=&\sin \theta _WW_{3L\mu }+\cos \theta _WB_\mu  \nonumber \\
Z_\mu &=&\cos \theta _WW_{3L\mu }-\sin \theta _WB_\mu  \nonumber \\
Z_{1\mu } &=&\frac g{g_1}W_{3R\mu }-\frac{g_X}{g_1}V_\mu  \nonumber \\
Z_{2\mu } &=&B_{1\mu }-\frac{g_X}{g_1}V_\mu  \nonumber \\
Z_\mu ^{\prime } &=&Z_{1\mu }-Z_{2\mu }=\frac g{g_1}W_{3R\mu }-B_{1\mu } 
\nonumber \\
Z_\mu ^{\prime \prime } &=&Z_{1\mu }+Z_{2\mu }=\frac g{g_1}W_{3R\mu
}+B_{1\mu }-2\frac{g_X}{g_1}V_\mu  \label{e26a}
\end{eqnarray}
The first stage of symmetry breaking is accomplished by a 15-plet of Higgs
scalar $\Phi $: 
\begin{equation}
\left\langle \Phi \right\rangle =\frac V2diag\left( 1,1,-1,-1\right)
\label{e27a}
\end{equation}
The second and third stages of symmetry breaking are accomplished by
introducing 4 Higgs scalars $\phi _1$, $\phi _2$, $\phi _3$, and $\phi _4$
belonging to fundamental representation of $SU_L\left( 4\right) $ having $%
Y_X=-2,0,0$ and 2 respectively, with the following expectation values 
\begin{eqnarray}
\left\langle \phi _1\right\rangle &=&v_1\left( 
\begin{array}{c}
1 \\ 
0 \\ 
0 \\ 
0
\end{array}
\right)  \nonumber \\
\left\langle \phi _2\right\rangle &=&v_2\left( 
\begin{array}{c}
0 \\ 
1 \\ 
0 \\ 
0
\end{array}
\right)  \nonumber \\
\left\langle \phi _3\right\rangle &=&v_3\left( 
\begin{array}{c}
0 \\ 
0 \\ 
1 \\ 
0
\end{array}
\right)  \nonumber \\
\left\langle \phi _4\right\rangle &=&v_4\left( 
\begin{array}{c}
0 \\ 
0 \\ 
0 \\ 
1
\end{array}
\right)  \label{e28a}
\end{eqnarray}
With the symmetry breaking pattern discussed above the mass Lagrangian for
vector bosons is given by 
\begin{eqnarray}
&&\left. {\cal L}_{mass}=\frac 14g_2^2V^2\left[ 2\bar{X}_1X_1+2\bar{X}_2X_2+2%
\bar{Y}_1Y_1+2\bar{Y}_2Y_2\right] \right.  \nonumber \\
&&+\frac 14g_2^2v_R^2\left[ 2\left( 1+\frac{{\cal K}^2+{\cal K}^{\prime 2}}{%
v_R^2}\right) W_R^{+}W_R^{-}+2\sin ^2\beta \left( \bar{Y}_1Y_1+\bar{Y}%
_2Y_2\right) +2\left( \cos ^2\beta +\frac{{\cal K}_R^2}{v_R^2}\right) \left( 
\bar{X}_1X_1+\bar{X}_2X_2\right) \right.  \nonumber \\
&&\left. +\frac{g_1^2}{g_2^2}\left( \left( \cos ^2\beta +\frac{{\cal K}_R^2}{%
v_R^2}\right) Z^{\prime 2}+\sin ^2\beta Z^{\prime \prime 2}\right) \right] 
\nonumber \\
&&+\frac 14g_2^2v_L^2\left\{ 2\left( 1+\frac{{\cal K}^2+{\cal K}^{\prime 2}}{%
v_L^2}\right) W_L^{+}W_L^{-}+2\frac{{\cal K}^2+{\cal K}^{\prime 2}}{v_R^2}%
W_R^{+}W_R^{-}\right.  \nonumber \\
&&+2\left( \sin ^2\alpha +\frac{{\cal K}^2+{\cal K}^{\prime 2}}{v_L^2}%
\right) \bar{X}_1X_1+2\left( \cos ^2\alpha +\frac{{\cal K}^2+{\cal K}%
^{\prime 2}}{v_L^2}\right) \bar{Y}_2Y_2+2\left( \sin ^2\alpha +\frac{4{\cal K%
}^2}{v_L^2}\right) \bar{Y}_1Y_1+2\left( \cos ^2\alpha +\frac{4{\cal K}^2}{%
v_L^2}\right) \bar{X}_2X_2  \nonumber \\
&&+\left( \sin ^2\alpha +\frac{{\cal K}^2+{\cal K}^{\prime 2}}{v_L^2}\right)
\left[ \frac Z{\cos \theta _W}+\frac{g_1}{g_2}\left( \tan ^2\theta
_W-1\right) +\frac 12\frac{g_1}{g_2}\tan ^2\theta _W\frac{g^2}{g_X^2}\left(
Z^{\prime }-Z^{\prime \prime }\right) \right] ^2  \nonumber \\
&&+\cos ^2\alpha \left[ \frac Z{\cos \theta _W}+\frac{g_1}{g_2}\tan ^2\theta
_W\left( Z^{\prime }+\frac 12\frac{g_2^2}{g_X^2}\left( Z^{\prime }-Z^{\prime
\prime }\right) \right) \right] ^2\left. +\frac{2{\cal KK}^{\prime }}{v_L^2}%
\left[ 2W_L^{+}W_R^{-}+2W_L^{-}W_R^{+}+2\bar{X}_1Y_2+2X_1\bar{Y}_2\right]
\right\}  \nonumber \\
&&  \label{e29}
\end{eqnarray}
where we have put $v_3=v_R\cos \beta $, $v_4=v_R\sin \beta $ and $%
v_1=v_L\sin \alpha $, $v_2=v_L\cos \alpha $. The following remarks are in
order

\begin{enumerate}
\item  For the symmetry breaking pattern which we have envsiged viz 
\begin{eqnarray*}
\left. SU_L\left( 4\right) \times U_X\left( 1\right) \right. 
\begin{tabular}[t]{c}
$\rightarrow $ \\ 
$V^2$%
\end{tabular}
SU_L\left( 2\right) \times SU_R\left( 2\right) \times U_1\left( 1\right)
\times U_X\left( 1\right)  \\
\,\,\,\,\,\,\,\,\,\,\,\,\,\,\,\,\,\,\,\,\,\,\,\,\,\,\,\,\,\,\,\,
\begin{tabular}[t]{c}
$\rightarrow $ \\ 
$v_R^2$%
\end{tabular}
SU_L\left( 2\right) \times U_1\left( 1\right) 
\begin{tabular}[t]{c}
$\rightarrow $ \\ 
$v_L^2$%
\end{tabular}
U_{em} \\
\left. V^2\gg v_R^2\gg v_L^2\right. 
\end{eqnarray*}

\item  For the symmetry breaking pattern in two stages, viz 
\[
SU_L\left( 4\right) \times U_X\left( 1\right) 
\begin{tabular}[t]{c}
$\rightarrow $ \\ 
$v_R^2$%
\end{tabular}
SU_L\left( 2\right) \times U_1\left( 1\right) 
\begin{tabular}[t]{c}
$\rightarrow $ \\ 
$v_L^2$%
\end{tabular}
U_{em}
\]
There is no need to introduce, 15-plet Higgs multiplet. This is an
attractive possibility

\item  In the decoupling limit, i.e. $g_X\rightarrow \infty $, $V_\mu $ is
decoupled and the minimum Higgs required for the symmetry breaking are Higgs
multiplets with $Y_X=0$ viz $\Phi $, $\phi _2$ and $\phi _3$. There are then
two neutral bosons $Z$ and $Z^{\prime }$. It effectively goes over to $%
SU_L\left( 4\right) $.

\item  The ratio between the vacuum expectation values $v_R$ and $v_L$ is
given by 
\[
\frac{v_R}{v_L}\sim \frac{m_X}{m_Z}\approx
33,\,\,\,\,\,\,for\,\,\,\,\,m_X=3\,\,\,TeV
\]
Using $v_L=264$ GeV, 
\[
v_R=33v_L=8.7\,\,\,\,\,\,TeV
\]

\item  In Eq. (36), the mass terms ${\cal K}_R$, ${\cal K}$, ${\cal K}%
^{\prime }$ arises from the vacuum expectation values of Higgs multiplets $%
S_{ij}$ and $S_{ij}^{\prime }$ (sec. III), which are required to give masses
to leptons and large Majorana mass to neutrinos $N_l$. We will assume ${\cal %
K}$, ${\cal K}^{\prime }\ll v_L$, so that their contributions to vector
bosons mass matrix can be neglected.The mixing between $W_L$ and $W_R$ is
determined by an angle 
\[
\epsilon \sim \frac{{\cal KK}^{\prime }}{v_R^2}=\frac{{\cal KK}^{\prime }}{%
v_L^2}\frac{v_L^2}{v_R^2}
\]
which is negligiblly small. In the limit of zero neutrino mass, ${\cal K}%
^{\prime }=0$, $\epsilon =0$ i.e. left-handed and right-handed currents are
completely decoupled from each other.
\end{enumerate}

\section{Lepton sector of the group $SU_L\left( 4\right) \times U_X\left(
1\right) $}

For leptons, since $Y_X$ is zero; the gauge invariant Lagrangian is similar
to that given in Ref. \cite{r09}; 
\begin{eqnarray}
{\cal L}_l &=&\bar{F}_\ell i\gamma ^\mu \left[ \partial _\mu +i\frac g2\vec{%
\lambda}\cdot \vec{W}_\mu \right] F_\ell  \nonumber \\
&\rightarrow &-\frac g{\sqrt{2}}\left[ \bar{\nu}_{\ell L}\gamma ^\mu \ell
_LW_{L\mu }^{-}+h.c.+\bar{N}_{\ell R}^{\prime }\gamma ^\mu \ell _RW_{R\mu
}^{-}+h.c.\right]  \nonumber \\
&&-\frac g2\left[ \bar{\ell}_L\gamma ^\mu \ell _L\left( W_{3L\mu }+\frac 1{%
\sqrt{2}}B_{1\mu }\right) +\bar{\nu}_{\ell L}\gamma ^\mu \nu _{\ell L}\left(
-W_{3L\mu }+\frac 1{\sqrt{2}}B_{1\mu }\right) \right]  \nonumber \\
&&+\bar{N}_{\ell R}^{\prime }\gamma ^\mu N_{\ell R}^{\prime }\left( W_{3R\mu
}-\frac 1{\sqrt{2}}B_{1\mu }\right) -\bar{\ell}_R\gamma ^\mu \ell _R\left(
W_{3R\mu }+\frac 1{\sqrt{2}}B_{1\mu }\right)  \nonumber \\
&&-\frac g{\sqrt{2}}\left[ -\bar{\ell}_R^c\gamma ^\mu N_{\ell L}^{\prime
}X_{1\mu }^{-}-\bar{\nu}_{\ell R}^c\gamma ^\mu N_{\ell L}X_{2\mu }^0-\bar{%
\ell}_R^c\gamma ^\mu \ell _RY_{1\mu }^{--}+\bar{\nu}_{\ell R}^c\gamma ^\mu
\ell _RY_{2\mu }^{-}+h.c.\right]  \nonumber \\
&=&-\frac{g_2}{\sqrt{2}}\left[ \bar{\nu}_{\ell L}\gamma ^\mu \ell _LW_{L\mu
}^{-}+h.c.+\bar{N}_{\ell R}^{\prime }\gamma ^\mu \ell _RW_{R\mu
}^{-}+h.c.\right]  \nonumber \\
&&-g_2\sin \theta _W\left( -\bar{\ell}\gamma ^\mu \ell \right) A_\mu -\frac{%
g_2}{2\cos \theta _W}\left[ \left( \bar{\nu}_{\ell L}\gamma ^\mu \nu _{\ell
L}-\bar{\ell}_L\gamma ^\mu \ell _L\right) +2\sin ^2\theta _W\bar{\ell}\gamma
^\mu \ell \right] Z_\mu  \nonumber \\
&&-\frac 12g_1\left[ \left( \bar{N}_{\ell R}\gamma ^\mu N_{\ell R}-\bar{\ell}%
_R\gamma ^\mu \ell _R\right) +\left( 1+\frac 12\frac{g_2^2}{g_X^2}\right)
\tan ^2\theta _W\left( \bar{\nu}_{\ell L}\gamma ^\mu \nu _{\ell L}-\bar{\ell}%
_L\gamma ^\mu \ell _L+2\bar{\ell}\gamma ^\mu \ell \right) \right] Z_\mu
^{\prime }  \nonumber \\
&&-\frac 12g_1\left[ -\frac 12\frac{g_2^2}{g_X^2}\tan ^2\theta _W\left( 2%
\bar{\ell}\gamma ^\mu \ell +\bar{\nu}_{\ell L}\gamma ^\mu \nu _{\ell L}-\bar{%
\ell}_L\gamma ^\mu \ell _L\right) \right] Z_\mu ^{\prime \prime }  \nonumber
\\
&&-\frac{g_2}{\sqrt{2}}\left[ -\bar{\ell}_R^c\gamma ^\mu N_{\ell R}^{\prime
}X_{1\mu }^{-}-\bar{\nu}_{\ell R}^c\gamma ^\mu N_{\ell R}^{\prime }X_{2\mu
}^0-\bar{\ell}_R^c\gamma ^\mu \ell _RY_{1\mu }^{--}+\bar{\nu}_{\ell
R}^c\gamma ^\mu \ell _RY_{2\mu }^{-}+h.c.\right]  \label{e33}
\end{eqnarray}
In the limit $g_X\rightarrow \infty $, it goes over to $SU_L\left( 4\right) $%
.

Note that $N_\ell ^{\prime }$ is not a mass eigenstate. It is related to
mass eigenstate $N_\ell $ as follows 
\begin{eqnarray}
N_\ell ^{\prime } &=&U_{\ell \ell ^{\prime }}N_\ell  \nonumber \\
\left( 
\begin{array}{c}
N_e^{\prime } \\ 
N_\mu ^{\prime } \\ 
N_\tau ^{\prime }
\end{array}
\right) &=&U\left( 
\begin{array}{c}
N_e \\ 
N_\mu \\ 
N_\tau
\end{array}
\right) ,\,\,\,UU^T=1  \label{e34}
\end{eqnarray}

To give masses to leptons, and large Majorana mass to the neutrino $N_\ell $%
, we introduce two Higgs multiplets $S_{ij}$ and $S_{ij}^{\prime }$
belonging to a symmetric representation 10 of $SU_L\left( 4\right) $. This
is done by giving the expectation values as follows: 
\begin{equation}
\left\langle S_{33}\right\rangle ={\cal K}_R,\,\,\,\,\,\,\,\,\,\,\,\,\,\,\,%
\,\,\,\,\left\langle S^{\prime }\right\rangle =\left( 
\begin{tabular}{cccc}
0 & 0 & 0 & $-{\cal K}$ \\ 
0 & 0 & ${\cal K}^{\prime }$ & 0 \\ 
0 & ${\cal K}^{\prime }$ & 0 & 0 \\ 
$-{\cal K}$ & 0 & 0 & 0
\end{tabular}
\right)  \label{e35}
\end{equation}
The scalars $S_{ij}$ and $S_{ij}^{\prime }$ have $Y_X=0$; thus they will be
coupled only to leptons and not to quarks. We shall assume that Yukawa
couplings of $S_{ij}$ and $S_{ij}^{\prime }$ are not very weak but have
normal strengths. Thus ${\cal K}$, ${\cal K}^{\prime }$, ${\cal K}_R$ are
very small compared to the vacuum expectation values of $\phi $'s viz ${\cal %
K}$, ${\cal K}^{\prime }\ll v_L$; ${\cal K}_R\ll v_R$. In this case, we may
neglect their contributions to masses of vector bosons.

The Yukawa couplings of these scalars to leptons are given by 
\begin{equation}
-f_\ell F_{\ell i}^TC^{-1}F_{\ell j}S_{ij}-f_\ell ^{\prime }F_{\ell
i}^TC^{-1}F_{\ell j}S_{ij}^{\prime }+h.c.  \label{e36}
\end{equation}
The lepton mass term is then given by 
\begin{equation}
{\frak m}_\ell =\left[ f_\ell ^{\prime }{\cal K}\bar{\ell}_L\ell _R+f_\ell
^{\prime }{\cal K}^{\prime }\bar{\nu}_{\ell L}N_{\ell R}^{\prime }-f_\ell 
{\cal K}_RN_{\ell R}^{\prime T}C^{-1}N_{\ell R}^{\prime }+h.c.\right]
\label{e37}
\end{equation}
From Eq. (\ref{e37}), the neutrino mass term can be written in the form: 
\begin{eqnarray}
{\frak m}_\nu &=&{\cal K}^{\prime }\left\{ f_e^{\prime }\bar{\nu}_{eL}\left(
U_{ee}N_{eR}+U_{e\mu }N_{\mu R}+U_{e\tau }N_{\tau R}\right) \right. +f_\mu
^{\prime }\bar{\nu}_{\mu L}\left( U_{\mu e}N_{eR}+U_{\mu \mu }N_{\mu
R}+U_{\mu \tau }N_{\tau R}\right)  \nonumber \\
&&\left. +f_\tau ^{\prime }\bar{\nu}_{\tau L}\left( U_{\tau e}N_{eR}+U_{\tau
\mu }N_{\tau R}+U_{\tau \tau }N_{\tau R}\right) \right\} +{\cal K}_R\left[
f_e\bar{N}_{eL}^cN_{eR}+f_\mu \bar{N}_{\mu L}^cN_{\mu R}+f_\tau \bar{N}%
_{\tau L}^cN_{\tau R}\right] +h.c.  \label{e38}
\end{eqnarray}

Note that we have obtained Eqs. (41) and (42) from Eq. (40) by using the
relation $\psi _R^T=-\bar{\psi}_L^cC$.

It is clear from Eq. (\ref{e38}) that the see-saw mechanism is a natural
consequence of this model. The flavor mixing for neutrinos is built in
feature of this model. The CKM type matrix $U$ can accomodate many features
of the neutrino oscillations which has been of considerable interest
recently. The lepton number violation occurs at a TeV scale. These processes
are mediated by vector bosons $X$, $Y$.

Some interesting lepton number violating processes that can occur at tree
level, through the exchange of $Y_2$ and $Y_1$ bosons are: 
\begin{eqnarray*}
\mu ^{-} &\rightarrow &\bar{\nu}_\mu +e^{-}+\nu
_e\,\,\,\,\,\,\,\,\,\,\,\,\,\,\,\,\,\,\,\,\,\,\,\Delta L_\mu =-2,\Delta L_e=2
\\
\mu ^{+}e^{-} &\rightarrow &\mu ^{-}e^{+} \\
\bar{\nu}_\mu e^{-} &\rightarrow &\bar{\nu}_e\mu ^{-}
\end{eqnarray*}
The experimental limit on the decay $\mu ^{-}\rightarrow \bar{\nu}_\mu
+e^{-}+\nu _e$%
\[
R=\frac{\Gamma \left( \mu ^{-}\rightarrow \bar{\nu}_\mu +e^{-}+\nu _e\right) 
}{\Gamma \left( \mu ^{-}\rightarrow all\right) }<1.2\times 10^{-2}; 
\]
Since 
\[
R\propto \left( \frac{m_W}{m_Y}\right) ^4, 
\]
the above limit implies 
\[
m_Y>3m_W 
\]
This implies a lower limit for the unification scale viz $m_X>3m_W$. Thus
the upper limit for the unification scale (5.8 TeV) is much above the
present experimental limit. Before we close this section, the following
comments are in order

The mass scale ${\cal K}$, ${\cal K}^{\prime }$ and ${\cal K}_R$ are
essentially determined by masses of charged leptons, light neutrinos and
heavy Majorana neutrinos. The see-saw mechanism gives the light neutrino
mass: 
\begin{eqnarray}
m_{\nu _l} &\sim &\frac{\left( f_l^{\prime }{\cal K}^{\prime }\right) ^2}{f_l%
{\cal K}_R}=\frac{m_l^2}{f_l{\cal K}_R}\left( \frac{{\cal K}^{\prime }}{%
{\cal K}}\right) ^2  \nonumber \\
f_l\left( \frac{{\cal K}}{{\cal K}^{\prime }}\right) ^2 &\sim &\frac{m_l^2}{%
m_{\nu _l}{\cal K}_R}
\end{eqnarray}
The atmospheric neutrino anomly implies $\Delta m_\nu ^2\sim 3\times 10^{-3}$
eV$^2$ and sets the scale of possible neutrino masses at $\geq 5\times
10^{-2}$ eV. On the other hand WMAP data [13] give $m_\nu <0.23$ eV. Hence
for $m_e=0.51$ MeV, we get 
\begin{eqnarray}
1 &<&f_e\left( \frac{{\cal K}^{\prime }}{{\cal K}}\right) ^2,  \nonumber \\
&<&5\,\,\,\,\,\,\,\,for\,\,\,\,\,\,\,\,{\cal K}_R=1\,\,\,\,TeV.
\end{eqnarray}
Hence there is no problem of heirarchy between the scales ${\cal K}$ and $%
{\cal K}^{\prime }$

\section{Quark sector of the group $SU_L\left( 4\right) \times U\left(
1\right) $}

For the first generation (for example), the interaction Lagrangian for
quarks is given by 
\begin{eqnarray}
{\cal L}_q &=&-\frac{g_2}{\sqrt{2}}\left[ \bar{u}_L\gamma ^\mu d_L^{\prime
}W_{L\mu }^{-}+\bar{D}_L\gamma ^\mu H_{dL}W_{L\mu }^{-}+h.c.\right] -g_2\sin
\theta _WJ_{em}^\mu A_\mu -\frac{g_2}{2\cos \theta _W}\left[ \bar{u}_L\gamma
^\mu u_L-\bar{d}_L\gamma ^\mu d_L-2\sin ^2\theta _WJ_{em}^\mu \right] Z_\mu 
\nonumber \\
&&-\frac 12g_1\left[ \left( \bar{D}_L\gamma ^\mu D_L-\bar{H}_{dL}\gamma ^\mu
H_{dL}\right) +\left( 1+\frac 12\frac{g_2^2}{g_X^2}\right) \tan ^2\theta
_W\left( \bar{u}_L\gamma ^\mu u_L-\bar{d}_L\gamma ^\mu d_L-2J_{em}^\mu
\right) \right] Z_\mu ^{\prime }  \nonumber \\
&&-\frac{g_1}2\left[ \frac 12\frac{g_2^2}{g_X^2}\tan ^2\theta _W\left( -\bar{%
d}_L\gamma ^\mu d_L+\bar{u}_L\gamma ^\mu u_L-2J_{em}^\mu \right) \right]
Z_\mu ^{\prime \prime }  \nonumber \\
&&-\frac{g_2}{\sqrt{2}}\left[ \bar{u}_L\gamma ^\mu D_LX_{1\mu }^{-}+\bar{d}%
_L^{\prime }\gamma ^\mu D_LX_{2\mu }^0+\bar{u}_L\gamma ^\mu H_{dL}Y_{1\mu
}^{--}+\bar{d}^{\prime }\gamma ^\mu H_{dL}Y_{2\mu }^{--}+h.c.\right]
\end{eqnarray}
where 
\begin{equation}
J_{em}^\mu =\frac 23\bar{u}\gamma ^\mu u-\frac 13\bar{d}\gamma ^\mu d-\frac 1%
3\bar{D}\gamma ^\mu D-\frac 43\bar{H}_d\gamma ^\mu H_d
\end{equation}
The quarks acquire masses from their Higgs 
\begin{eqnarray}
\phi _1 &=&\left( 
\begin{array}{c}
\phi _1^0 \\ 
\phi _1^{-} \\ 
\phi _1^{\prime -} \\ 
\phi _1^{--}
\end{array}
\right) \,\,\,\,\,\left\langle \phi _1^0\right\rangle =v_1=v_L\sin \alpha 
\nonumber \\
\phi _2 &=&\left( 
\begin{array}{c}
\phi _2^{+} \\ 
\phi _2^0 \\ 
\phi _2^{\prime 0} \\ 
\phi _2^{-}
\end{array}
\right) \,\,\,\,\,\left\langle \phi _2^0\right\rangle =v_2=v_L\cos \alpha 
\nonumber  \label{e30} \\
\phi _3 &=&\left( 
\begin{array}{c}
\phi _3^{+} \\ 
\phi _3^0 \\ 
\phi _3^{\prime 0} \\ 
\phi _3^{-}
\end{array}
\right) \,\,\,\,\,\left\langle \phi _3^{\prime 0}\right\rangle =v_3=v_R\cos
\beta  \nonumber \\
\phi _4 &=&\left( 
\begin{array}{c}
\phi _4^{++} \\ 
\phi _4^{+} \\ 
\phi _4^{\prime +} \\ 
\phi _4^0
\end{array}
\right) \,\,\,\,\,\left\langle \phi _4^0\right\rangle =v_4=v_R\sin \beta
\label{e31}
\end{eqnarray}

Let us write down the Yukawa couplings of scalars $\phi $'s with the quarks:
(here we depict the couplings for the first generation) 
\begin{equation}
{\cal L}_Y=f_u\bar{F}_d\phi _1u_R+f_d\bar{F}_d\phi _2d_R^{\prime }+f_d\bar{F}%
_d\phi _2D_R+f_D\bar{F}_d\phi _3D_R+f_D\bar{F}_d\phi _3d_R^{\prime }+f_{H_d}%
\bar{F}_d\phi _4H_{d_R}+h.c.
\end{equation}

Note that while the scalars $\phi _1$ and $\phi _4$ are coupled to $u_R$ and 
$H_{d_R}$ respectively; the scalars $\phi _2$ and $\phi _3$ are coupled to
both $d_R$ and $D_R$, giving off-diagonal mass terms for $d_R$ and $D_R$. In
order to avoid this, we require the Yukawa couplings to be invariant under a
discrete symmetry: under which 
\begin{eqnarray*}
\phi _1,\phi _2 &\rightarrow &\phi _1,\phi _2, \\
\phi _3,\phi _4 &\rightarrow &-\phi _3,-\phi _4, \\
F_d &\rightarrow &F_d,u_R,d_R\rightarrow u_R,d_R \\
D_R,H_{d_R} &\rightarrow &-D_R,-H_{d_R}
\end{eqnarray*}
In this way, the quarks acquire masses given below 
\begin{eqnarray}
m_u &=&f_uv_L\sin \alpha ,m_d=f_dv_L\cos \alpha  \nonumber \\
m_D &=&f_Dv_R\cos \beta ,m_{H_d}=f_{H_d}v_R\sin \beta
\end{eqnarray}
Thus, the quarks $D$ and $H_d$ are superheavy.

Now we have two types of Higgs coupled to down quarks. Thus we must address
the question of suppression of flavor changing neutral currents (FCNC). At
tree level, there are no FCNC. However, they may appear at loop level. FCNC\
can be eliminated by replacing $D$ and $S$ by $D^{\prime }$ and $S^{\prime }$%
, where $D^{\prime }$ and $S^{\prime }$ are related to mass eigenstates $D$
and $S$ with same Cabibbo rotation as for $d^{\prime }$ and $s^{\prime }$.
This is a reasonable assumption as $D$ and $S$ are replicas of $d$ and $s$
at a higher mass scale.

To conclude, The electroweak unification model based on the group $%
SU_L\left( 4\right) \times U_X\left( 1\right) $ has following features: In
the leptonic sector, it contains heavy right-handed neutrinos $N_e$, $N_\mu $
and $N_\tau $. The see-saw mechanism for the neutrino mass matrix is built
in the model. $\Delta L_\mu =\pm 2$, $\Delta L_e=\mp 2$ processes can occure
due to exchange of heavy bosons $X$ and $Y$. In the quark sector we have six
extra heavy quarks; three of them viz $\left( U,D,S\right) $ have charges $%
\left( 2/3,-1/3,-1/3\right) $ and are mirrors of $\left( u,d,s\right) $
quarks. However, these quarks would decay to light quarks by $\beta $-decay
with the emission of right-handed Majorana neutrino at tree level through
the exchange of $X$-bosons. A typical process is depicted below $\left(
\Delta L_e=2\right) $%
\[
D\rightarrow u+e^{-}+N_e 
\]
The superheavy quarks $\left( U,D,S\right) $ can form bound states with the
light quarks $u$, $d$, $s$, $c$ and $b$. Thus this model predict a replica
of existing hadrons at a scale of few hundred GeV to TeV. The superheavy
quarks $H_d$, $H_s$, and $T$, have exotic charges $\left(
-4/3,-4/3,5/3\right) $. If these quarks have masses greater than $Y$ bosons,
they would decay quickly to light quarks by a process of the form $%
H_d\rightarrow u+Y_1^{--}$. Even if they have masses below $Y$-bosons, they
are expected to decay quickly to light leptons by a typical process:
(neutrinoless $\beta $-decay; $\Delta L_e=2$) 
\[
H_d\rightarrow u+l^{-}+l^{-} 
\]
Thus $H_d$, $H_s$, and $T$ may not live long enough to form bound states
with light quarks.

Finally, from the interaction Lagrangians given in Eqs. (37) and (45), it is
clear that for $g_2^2/g_X^2\ll 1$, we get effectively interaction
Lagrangians of group $SU_L\left( 4\right) $. Thus for $g_2^2/g_X^2\ll 1$, it
is still possible to have unification at TeV scale.

In the end, it is interesting to note that any viable extension of the
standard model $\left( SU\left( 3\right) _c\times SU_L\left( 2\right) \times
U_Y\left( 1\right) \right) $ is not possible without extending the existing
elementary consituents of matter (viz six leptons and six quarks). In this
sense, the standard model is unique if there are no additional constituents
of matter.

Note added: After putting the paper on the arXiv, we were informed by Dr. R.
Foot that the group $SU_L\left( 4\right) $ originally considered by us \cite
{r09} has been extended earlier to $SU_L\left( 4\right) \times U\left(
1\right) $ \cite{r11,r12} to include fractionally charged quarks. These
extensions do not contain any reference to our earlier work \cite{r09}. The
motivation of our work is some what different, although there is overlap
with Ref. \cite{r11,r12}. Moreover, in our version, see-saw mechanism for
neutrino mass matrix is a natural consequence.

Acknowledgement: This work was supported in part by a\ grant from Pakistan
Council for Science and Technology.

\end{document}